\documentclass{ws-procs10x7}
\usepackage{balance}
\usepackage{epsfig}

\newcolumntype{d}[1]{D{.}{.}{#1}}

\newcommand{\Reepe}{\text{Re(}\epsilon'/\epsilon \rm{)}}

\newcommand{\kpmpipizpiz}{K^\pm \to \pi^\pm \pi^0 \pi^0}

\newcommand{\kpmpipig}{K^\pm \to \pi^\pm \pi^0  \gamma}

\newcommand{\xilampi}{\Xi^0 \to \Lambda \pi^0}

\newcommand{\xilamgam}{\Xi^0 \to \Lambda \gamma}

\newcommand{\xisiggam}{\Xi^0 \to \Sigma^0 \gamma}

\newcommand{\bdm}{\begin{displaymath}}
\newcommand{\edm}{\end{displaymath}}
\newcommand{\be}{\begin{equation}}
\newcommand{\ee}{\end{equation}}

\def\Journal#1#2#3#4{{\it #1} {\bf #2}, #3 (#4)}

\makeindex
\begin{document}

\title{NEW NA48 RESULTS ON KAON AND HYPERON DECAYS}

\author{R.~WANKE}

\address{Institut f\"ur Physik, Universit\"at Mainz, D-55099 Mainz, Germany \\$^*$E-mail: Rainer.Wanke@uni-mainz.de}


\twocolumn[\maketitle\abstract{
The NA48 Collaboration has obtained several new results on kaon and neutral hyperon decays.
For hyperons we present new, very precise measurements on the $\Xi^0$ lifetime and the decay asymmetries
of $\xilamgam$ and $\xisiggam$. On kaon decays, we report on the first observation of the
interference term between Inner Bremsstrahlung and Direct Emission in $\kpmpipig$ and on a new, accurate
measurement of the form factor slopes in the semileptonic decay $K_L \to \pi^\pm \mu^\mp \nu$.
}]

\section{Introduction}

The NA48 experiment is a fixed-target experiment located at the CERN SPS and
has been taken data from 1997 to 2004.
Its first phase, ending in 2001, was dedicated to the measurement
of direct CP violation in the neutral kaon system. It was employing
simultaneous beams of $K_L$ and $K_S$ meson decays for measuring the parameter
$\Reepe$ of direct CP violation in the decays $K_L/K_S \to \pi^+ \pi^-/\pi^0 \pi^0$.
Using the same data sample, also
several semileptonic and rare $K_L$ decays could be precisely measured.

The second phase of the experiment, dubbed NA48/1, has been taken data in the year 2002.
In this set-up, the $K_L$ beam was blocked, but the proton intensity on the $K_S$ target
was increased by a factor of 200 for measuring rare $K_S$ and neutral hyperon decays.

Finally, for the NA48/2 experiment in 2003 and 2004, the beam set-up was changed
again in order to simultaneously record $K^+$ and $K^-$ decays. 
Main purpose of this experiment was the search for direct CP violation in $K^\pm$ decays
to three pions, but also high statistics measurements of semileptonic, rare, and very rare charged
kaon decays.

Apart from minor modifications and upgrades, as e.g.\ the introduction of a beam spectrometer
for NA48/2, the detector was practically unchanged during the whole period.
A detailed description of the detector can be found elsewhere~\cite{bib:reepe}.
Presented here are results on $\Xi^0$ and kaon decays from all three phases of the experiment.

\section{$\Xi^0$ Decays}

\subsection{$\Xi^0$ Lifetime}

The $\Xi^0$ lifetime is an important input to other measurements, first of all
to the determination of $|V_{us}|$ via the rate of the $\Xi^0$ $\beta$-decay.
Its current precision is only about $3\%$ from measurements dating back to the 60's and 70's~\cite{bib:pdg06}.

\begin{figure}
\centerline{\epsfig{file=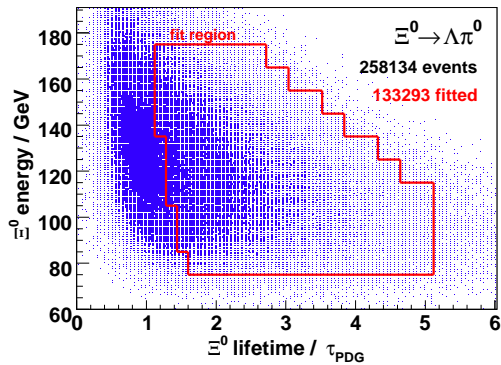,width=\linewidth}}
\caption{Distribution of energy versus proper lifetime for 
selected $\xilampi$ minimum bias events. Indicated by the red enclosure are the fit regions for the differnt energy bins.}
\label{fig:Xi0Etau}
\end{figure}

Using NA48/1 minimum bias data, we have selected about 260000 practically background-free
$\xilampi$ decays. A fit was done simultaneously in 10 bins of energy with
the 10 normalizations and the $\Xi^0$ lifetime as free parameters. The fit region in energy and lifetime is shown 
in Fig.~\ref{fig:Xi0Etau}. Care was taken to avoid detector resolution effects near the collimator.

As preliminary result we obtain
\begin{equation}
  \tau(\Xi^0) = (3.082 \pm 0.013_{\text{stat}} \pm 0.012_{\text{sys}}) \times 10^{-10} \; \text{s}.
\end{equation}

\begin{figure}
\centerline{\epsfig{file=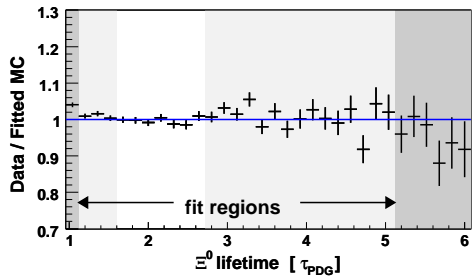,width=\linewidth}}
\caption{$\xilampi$ data divided by fitted MC. In the light grey region not all energy bins
         were included in the fit, the dark grey region was excluded from the fit.}
\label{fig:Xi0lifetimefit}
\end{figure}

The agreement between data and fitted MC is very good (Fig.~\ref{fig:Xi0lifetimefit}).
The systematics are dominated by uncertainties of the detector acceptance
and the calorimeter energy scale.
The result is about two standard deviations above the current PDG average~\cite{bib:pdg06} and five times more precise.

\subsection{$\Xi^0$ Decay Asymmetries}

\begin{figure}
\centerline{\epsfig{file=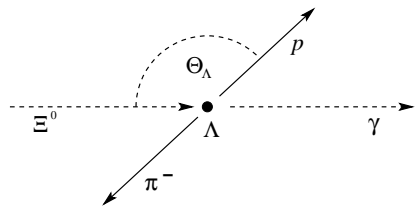,width=0.8\linewidth}}
\caption{Definition of the angle $\Theta$ between the proton and the incoming $\Xi^0$ in the Lambda rest frame.}
\label{fig:Xi0Lamgamsketch}
\end{figure}

Up to this day, weak radiative hyperon decays as $\xilamgam$ and $\xisiggam$ are still barely understood. 
Several competing theoretical models exist, which give very different predictions.
An excellent experimental parameter to distinguish between models is the decay asymmetry $\alpha$
of these decays. It is defined as 
\begin{equation}
  \frac{d N}{d \cos \Theta} = N_0 ( 1 + \alpha \, |\vec{P}| \, \cos \Theta),
\end{equation}
where $\Theta$ is the direction of the daughter baryon with respect to the polarization $\vec{P}$
of the mother in the mother rest frame.
For e.g.\ $\xilamgam$, the decay asymmetry can then be measured by looking at the 
angle between the incoming $\Xi^0$ and the outgoing proton from the subsequent $\Lambda \to p \pi^-$ decay
in the $\Lambda$ rest frame (see Fig.~\ref{fig:Xi0Lamgamsketch})
Using this method, the measurement is independent of the unknown initial $\Xi^0$ polarization.

\begin{figure}
  \centerline{\epsfig{file=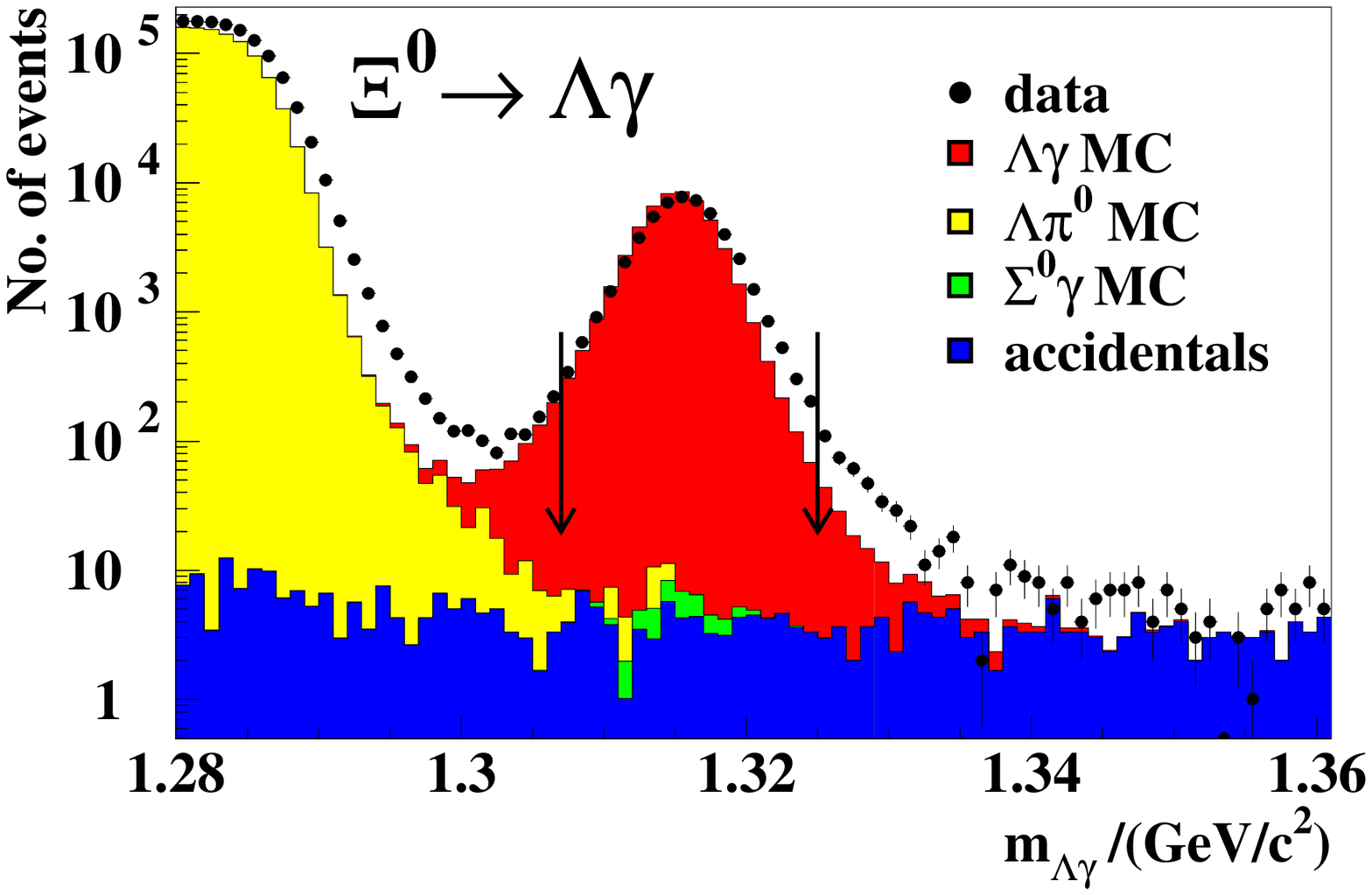,width=\linewidth}}
  \centerline{\epsfig{file=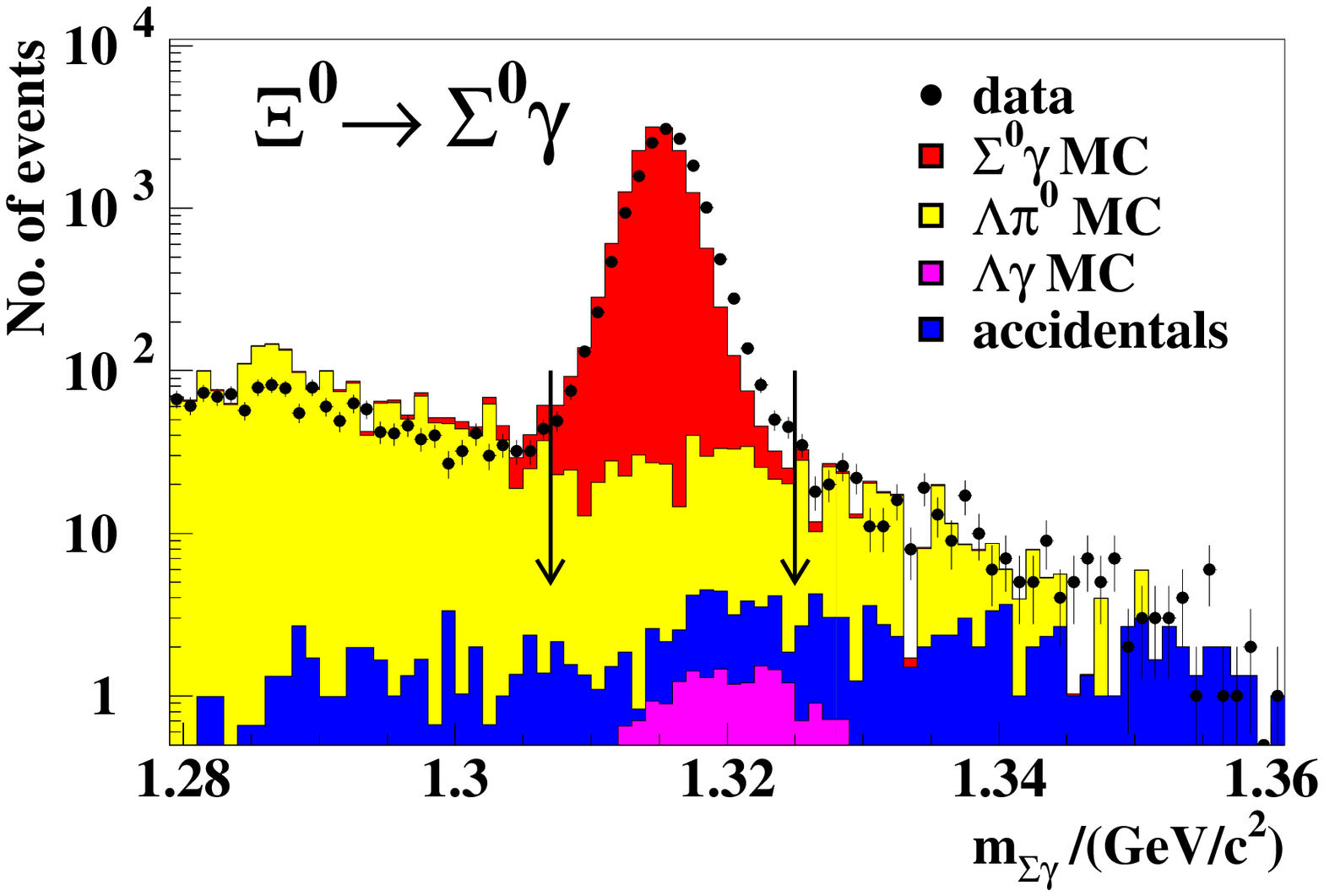,width=\linewidth}}
  \caption{$\xilamgam$ (top) and $\xisiggam$ (bottom) signal together with MC expectations for signal and backgrounds.}
  \label{fig:Xi0signals}
\end{figure}

The NA48/1 experiment has seen a total $\Xi^0$ flux of about $2.4 \times 10^9$ decays in the fiducial volume.
From this data set, 48314 $\xilamgam$ and 13068 $\xisiggam$ candidates have been collected (Fig.~\ref{fig:Xi0signals}).
The background contributions are $0.8\%$ for $\xilamgam$ and about $3\%$ for $\xisiggam$, respectively.

\begin{figure}
  \centerline{\epsfig{file=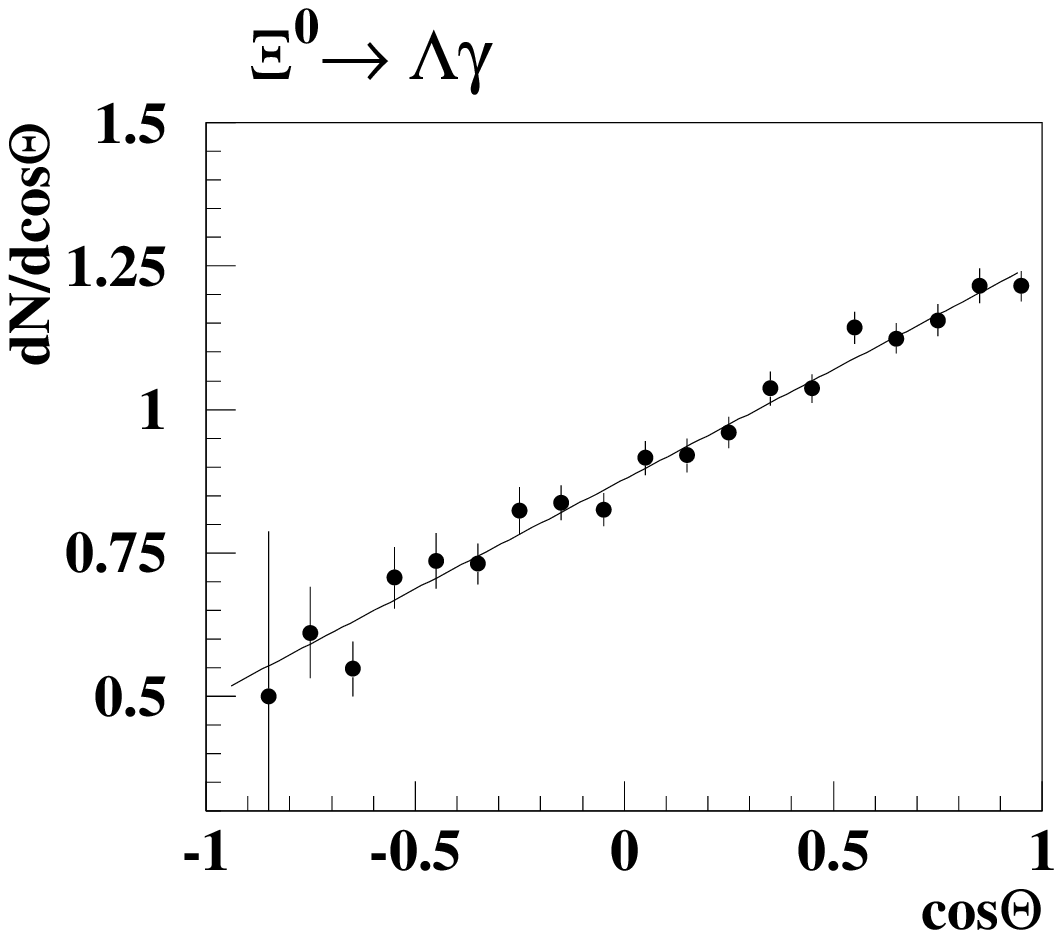,width=0.86\linewidth}}
  \vspace*{-6mm}
  \centerline{\epsfig{file=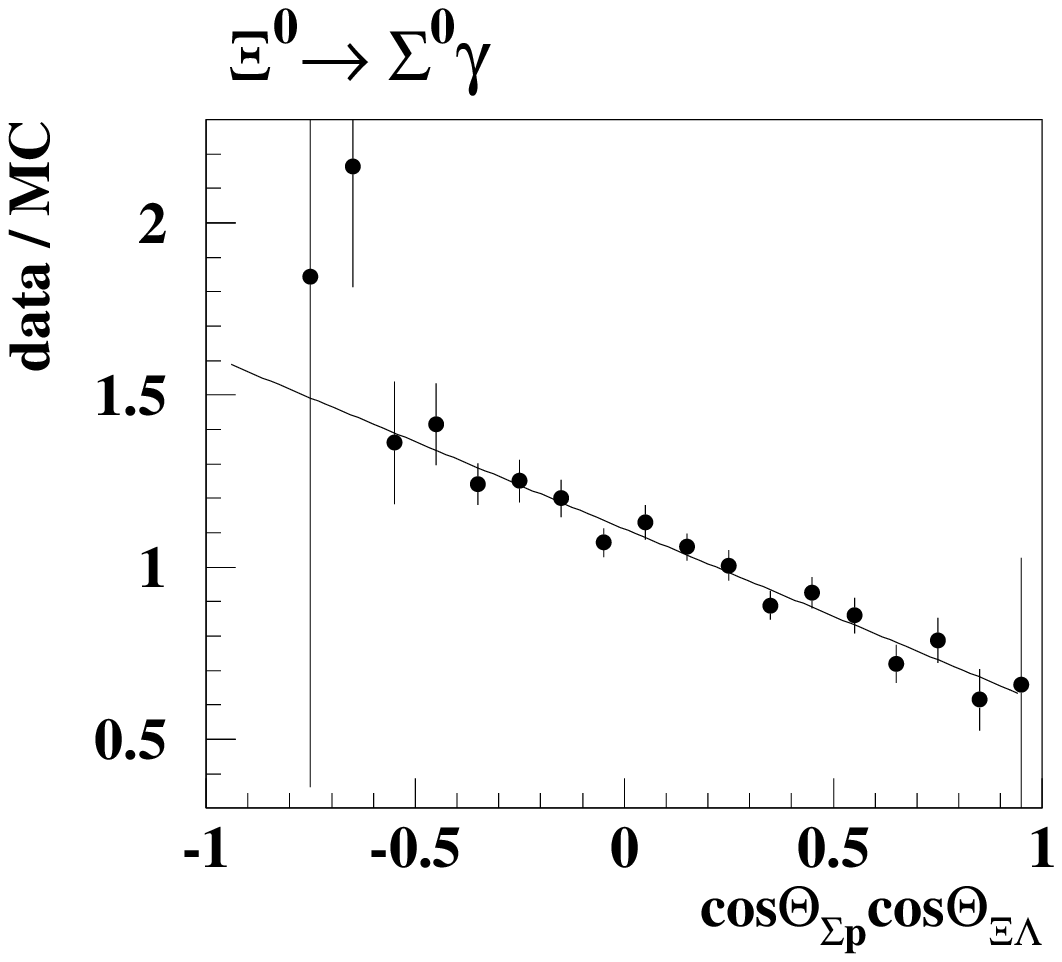,width=0.86\linewidth}}
  \caption{Fits of the decay asymmetries in $\xilamgam$ (top) and $\xisiggam$ (bottom).}
  \label{fig:Xi0asymmetries}
\end{figure}

Using these data, fits to the decay asymmetries have been performed. In case of $\xisiggam$, where
we have the subsequent decay $\Sigma^0 \to \Lambda \gamma$, the product 
$\cos \Theta_{\Xi \to \Sigma \gamma} \cdot \cos \Theta_{\Sigma \to \Lambda \gamma}$
has to be used for the fit.
Both fits show the expected linear behaviour on the angular parameters (Fig.~\ref {fig:Xi0asymmetries})

After correcting for the well-known asymmetry of $\Lambda \to p \pi^-$, we obtain
\begin{eqnarray}
  \alpha_{\Xi^0 \to \Lambda \gamma}  & = & -0.684 \pm 0.020 \pm 0.061, \\
  \alpha_{\Xi^0 \to \Sigma^0 \gamma} & = & -0.682 \pm 0.031 \pm 0.065,
\end{eqnarray}
where the first error is statistical and the second systematic.
These values agree with previous measurements by NA48 on $\xilamgam$~\cite{bib:Xi0Lamgam_NA48}
and KTeV on $\xisiggam$~\cite{bib:Xi0Siggam_KTeV}, but are much more precise.
In particular the result on $\xilamgam$ is of high theoretical interest, as it confirms the large
negative value of the decay asymmetry, which is difficult to accommodate for quark and vector meson dominance models.

\section{Kaon Decays}

\subsection{$\kpmpipig$ Decays}

The decay $\kpmpipig$ is dominated by Inner Bremsstrahlung (IB) from the charged pion.
However, there exist small components of Direct Emission (DE), which are sensitive to electric and magnetic dipole amplitudes, and also from the
interference (INT) between IB and DE, which only depends on the electric dipole. These components are of large interest within the framework 
of chiral perturbation theory. Previous experiments have found the DE component, but could only set limits 
on the interference term~\cite{bib:pdg06}.

\begin{figure}
\centerline{\epsfig{file=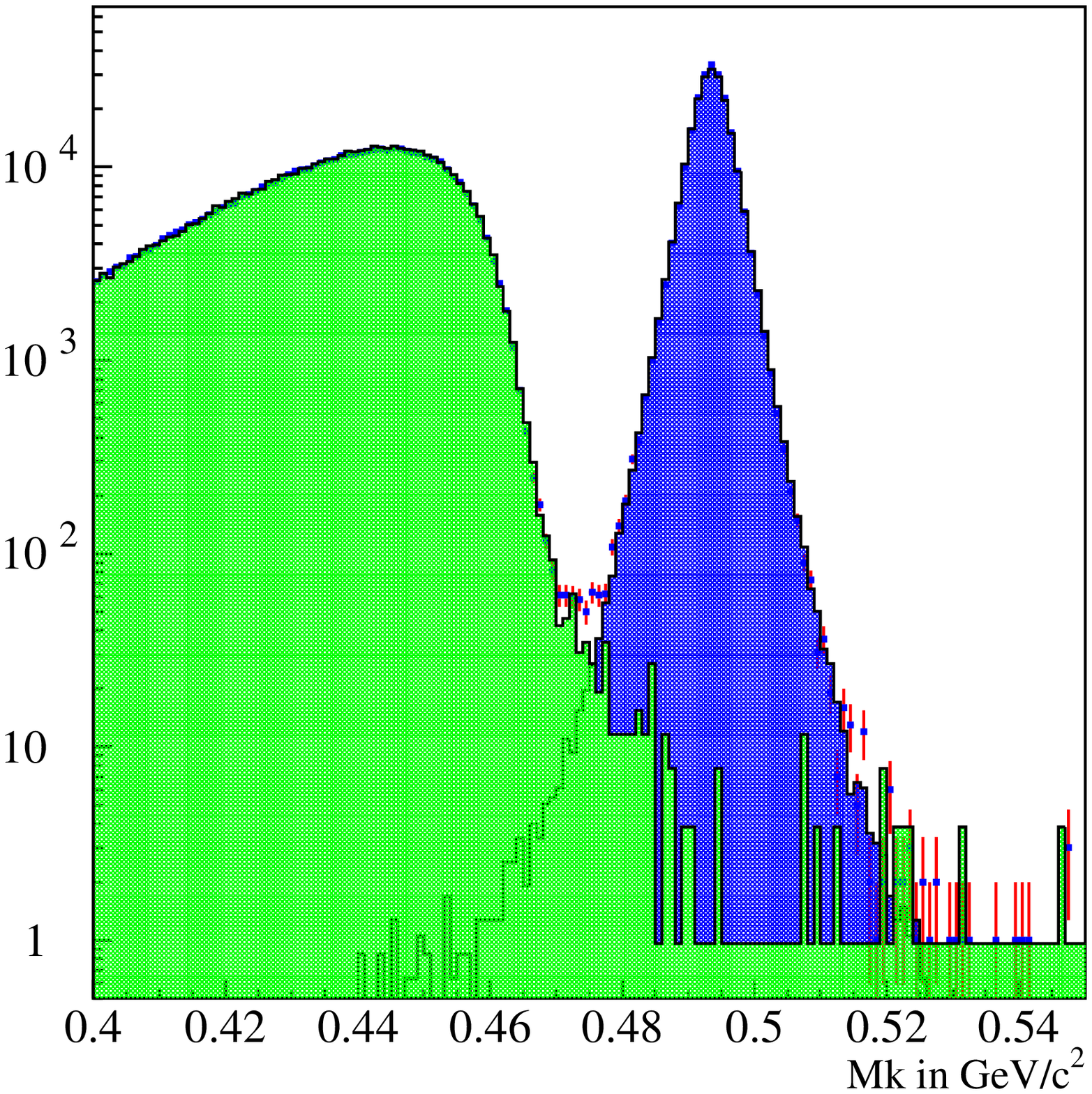,width=0.9\linewidth}}
\caption{Invariant $\pi^\pm \pi^0 \gamma$ mass for selected $\kpmpipig$ events.
         Also shown are the MC simulations of the signal and the background from $\kpmpipizpiz$.}
\label{fig:KpipigMass}
\end{figure}

The NA48/2 experiment has collected about one million $\kpmpipig$ decays with almost negligible background.
So far we have used a sub-sample of about 210000 events of these data for the analysis. The events have been selected
in the range $0 < T^\star_\pi < 80$~MeV of the kinematic energy of the $\pi$ in the kaon rest frame.
The invariant mass distribution of these events is shown
in Fig.~\ref{fig:KpipigMass}.

\begin{figure}
\centerline{\epsfig{file=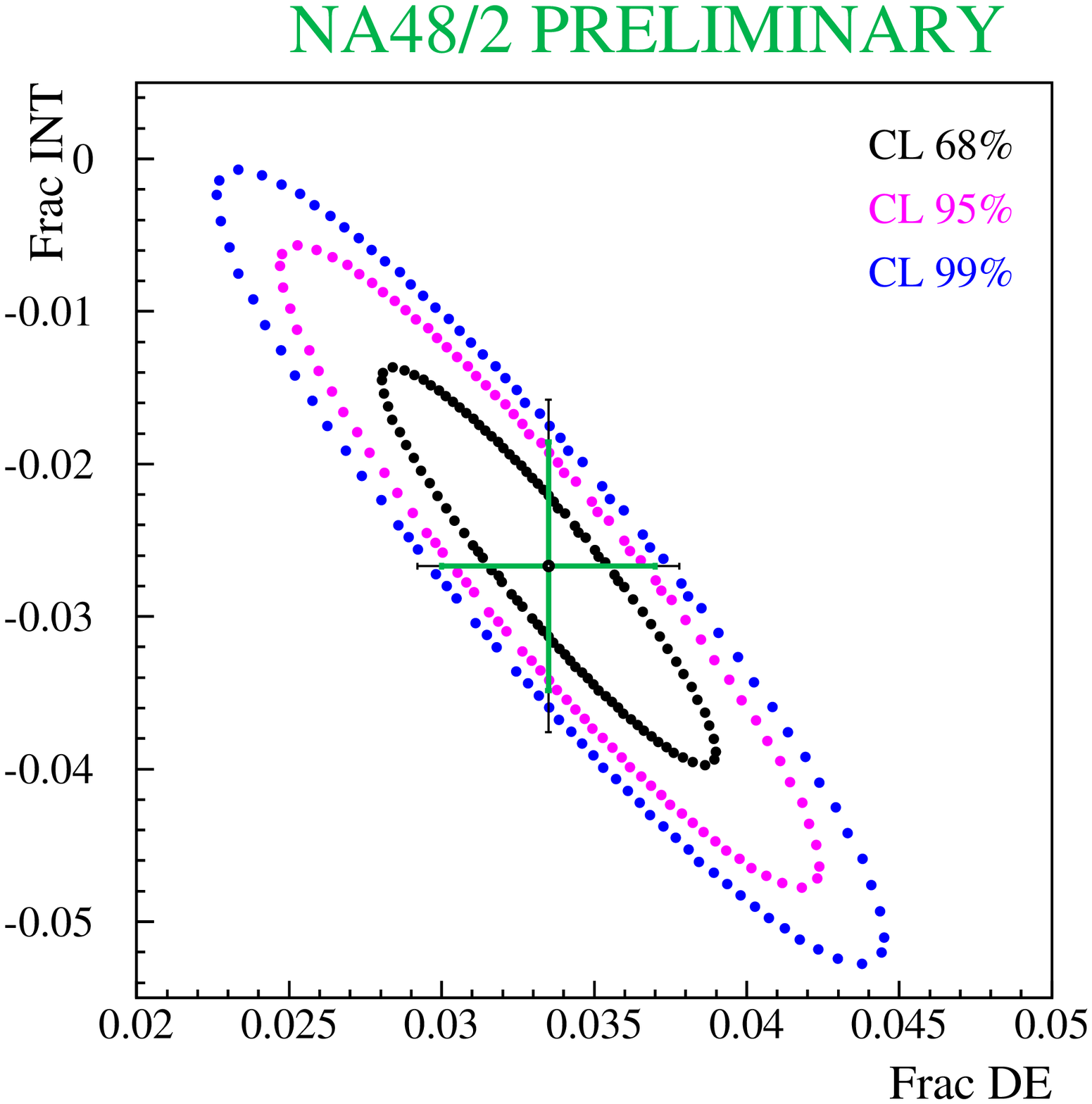,width=0.93\linewidth}}
\caption{Confidence regions for the fractions of DE and interference of the $\kpmpipig$ amplitude.}
\label{fig:KpipigResult}
\end{figure}

The lorentz-invariant variable sensitive to the non-IB amplitudes is $W \equiv (p_K p_\gamma)(p_\pi p_\gamma)/(m_K m_\pi)^2$.
Fitting the $W$ distribution with the three components IB, DE, and INT, we obtain for the fractions of DE and INT
\begin{eqnarray}
  \text{Frac(DE)}  & = & (3.35  \pm 0.35 \pm 0.25)\%, \\
  \text{Frac(INT)} & = & (-2.67 \pm 0.81 \pm 0.73)\%.
\end{eqnarray}
The first error is statistical, the second systematic.
The systematics are dominated by the trigger efficiency.
The statistical correlation between the two components is -0.92 (see Fig.~\ref{fig:KpipigResult}).
This is the first evidence for a non-zero interference term in this decay.

For comparison with previous results we have also performed a single parameter fit, setting INT to zero, and extrapolated
to the region $55 < T^\star_\pi < 90$~MeV. The result then becomes
$\text{Frac(DE)}_{55 < T_\pi^\star < 90 \: \text{MeV}}^{\text{INT} = 0} =  (0.85  \pm 0.05 \pm 0.02)\%$.

\subsection{$K^0_{\mu3}$ Form Factors}

\begin{figure}
\centerline{\epsfig{file=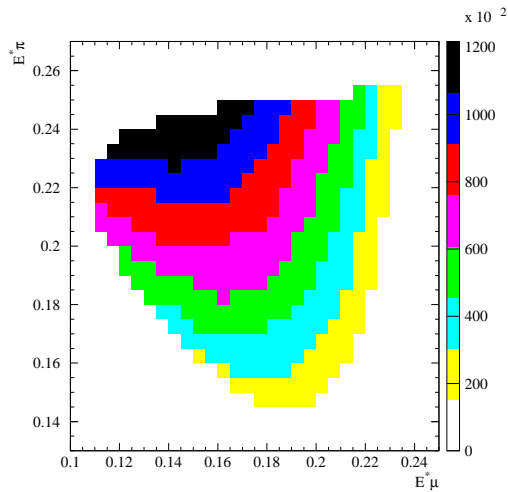,width=\linewidth}}
\caption{Dalitz plot distribution of selected $K^0_{\mu3}$ events.}
\label{fig:Kmu3Dalitzplot}
\end{figure}

In a special two-day $K_L$ run in 1999, taken with a minimum bias trigger, we have found
2.6~million $K^0_{L\,\mu3}$ decays. The Dalitz plot of these decays is shown in Fig.~\ref{fig:Kmu3Dalitzplot}.

A fit to the Dalitz plot yields for the slopes $\lambda_+$ and $\lambda_0$ of the
vector and scalar form factor, resp., 
\begin{eqnarray}
  \lambda_+ & = & 0.0260 \pm 0.0007_{\text{stat}} \pm 0.0010_{\text{syst}}, \\
  \lambda_0 & = & 0.0120 \pm 0.0008_{\text{stat}} \pm 0.0015_{\text{syst}}.
\end{eqnarray}
Both these values disagree with the former PDG04 values~\cite{bib:pdg04}. The agreement is better with a recent
KTeV analysis~\cite{bib:KTeV_Kmu3ff}, however, at least for $\lambda_0$ not perfect either.
The result is still preliminary. A more detailed analysis, using also quadratic slopes and pole
form factors will be published soon.



\begin{thebibliography}{99}

\bibitem{bib:reepe}  A.~Lai {\it et al.}, 
                     \Journal{Eur.~Phys.~J.}{C \bf 22}{231}{2001}.

\bibitem{bib:pdg06}  W.-M.~Yao {\it et al.}, 
                     \Journal{Jour.~of~Phys.}{G \bf 33}{1}{2006}.

\bibitem{bib:Xi0Lamgam_NA48}  A.~Lai {\it et al.}, 
                              \Journal{Phys.~Lett.}{B \bf 584}{251}{2004}.

\bibitem{bib:Xi0Siggam_KTeV}  A.~Alavi-Harati {\it et al.}, 
                              \Journal{Phys.~Rev.~Lett.}{\bf 86}{3239}{2001}.

\bibitem{bib:pdg04}  S.~Eidelman {\it et al.}, 
                     \Journal{Phys.~Lett.}{B \bf 592}{1}{2004}.

\bibitem{bib:KTeV_Kmu3ff}  T. Alexopoulos {\it et al.},
                           \Journal{Phys.~Rev.}{D \bf 70}{092007}{2004}.


\end{thebibliography}
\end{document}